\documentclass[aps,jcp,twocolumn,showpacs,superscriptaddress,groupedaddress]{revtex4}
\usepackage{latexsym}
\usepackage{amsfonts,amsbsy,eucal,dsfont}
\usepackage{amssymb}
\usepackage{amsmath}
\usepackage{dsfont}
\usepackage{graphicx}
\usepackage{float}
\usepackage{graphicx}
\usepackage{bm}
\usepackage{natbib}
\usepackage{color}

\renewcommand{\figurename}{Fig.}

\renewcommand{\refname}{Ref.}
\newcommand{\refsname}{Refs.}
\newcommand{\expressionname}{Eq.}
\newcommand{\expressionsname}{Eqs.}
\newcommand{\sectionname}{Sec.}

\renewcommand{\tablename}{Tab.}

\begin{document}

\title{Coupling between bulk- and surface chemistry in suspensions of charged colloids}

\author{M. Heinen}
\email[]{mheinen@thphy.uni-duesseldorf.de}
\affiliation{Institut f\"{u}r Theoretische Physik II, Weiche Materie, Heinrich-Heine-Universit\"{a}t D\"{u}sseldorf, 40225 D\"{u}sseldorf, Germany}
\author{T. Palberg}
\affiliation{Institut f\"{u}r Physik, Johannes Gutenberg Universit\"{a}t Mainz, 55128 Mainz, Germany}
\author{H. L\"{o}wen}
\affiliation{Institut f\"{u}r Theoretische Physik II, Weiche Materie, Heinrich-Heine-Universit\"{a}t D\"{u}sseldorf, 40225 D\"{u}sseldorf, Germany}

\date{\today}

\begin{abstract}
The ionic composition and pair correlations in fluid phases of realistically salt-free charged colloidal sphere suspensions are calculated in the primitive model.
We obtain the number densities of all ionic species in suspension, including low-molecular weight microions, and 
colloidal macroions with acidic surface groups, from a self-consistent solution of a coupled physicochemical set of nonlinear algebraic
equations and non-mean-field liquid integral equations. Here, we study suspensions of colloidal spheres with sulfonate or silanol surface groups, suspended in demineralized water
that is saturated with carbon dioxide under standard atmosphere.
The only input required for our theoretical scheme are the acidic dissociation constants $pK_a$, and effective sphere diameters of all involved ions.
Our method allows for an \textit{ab initio} calculation of colloidal bare and effective charges, at high numerical efficiency.  
\end{abstract}

\pacs{82.70.Dd. 
      82.70.Kj, 
      61.20.-p, 61.25.-f, 
      78.30.cd, 
      }
\maketitle

\section{Introduction}
 
Predicting the structural correlations in suspensions of charged colloidal particles without
any fitting parameters still represents a formidable challenge of statistical physics.
This is mainly due to two reasons: first, the Coulomb interactions are long-ranged and
there are nontrivial correlations between the colloidal macroions and between the microions
which require an extension of standard mean-field theories of linear screening
\cite{Hansen2000, Levin2002, Messina2009}.
Second, the (bare) charge of the colloidal particles in suspension is not known \textit{a priori}, but underlies the chemical \textit{charge regulation}
process, with the dissociation degree of ionizable colloidal surface groups depending on the amount of added electrolyte ions and on the colloidal concentration.
The resulting colloidal bare charge largely differs from the titration charge, \textit{i.e.}, the maximal
possible charge for a colloidal particle with fully dissociated acidic surface groups \cite{Wette2001, Wette2002}.

In addition, presence of microions with non-mean-field like distributions in narrow diffusive layers about the colloidal particle's surfaces
\cite{Trizac2003, Torres2008, Pianegonda2007, McPhie2008, Rojas-Ochoa2008, Colla2009, Castaneda-Priego2012, Heinen2013}
causes that the \textit{effective} electrostatic interaction is further reduced.
For instance, in a one component macroion fluid model, where the microion degrees of freedom are integrated out \cite{Verwey_Overbeek1948},
it is an effective colloidal charge that dictates the pair-correlations among colloidal particles.
Typically, the colloidal (effective) charge is treated as a fit parameter.
An example is the fitting of a Debye-H\"{u}ckel potential to the far-field numerical non-linear Poisson-Boltzmann solution \cite{Alexander1984}.
The so-determined type of effective charge is also known as renormalized charge.
In experimental analysis, an effective charge is commonly used 
in describing colloidal static structure factors or radial distribution functions
\cite{Nagele1996, Haertl1988, PhilipseVrij1988, Royall2006, Gapinski2009, Heinen2011, Holmqvist2012, Heinen2012, Westermeier2012, Gruijthuijsen2013}.
Also the phase behavior \cite{Wette2010} and the elastic properties in the solid state \cite{Wette2002, Wette2003}
can be interpreted in terms of a Debye-H\"{u}ckel potential, based on a fitted effective charge, and,
furthermore, colloidal effective charges determine the suspension's electro-kinetic properties \cite{Wette2001, Palberg2013, Medebach2007}.
Conductivity measurements, in particular, access the number of uncondensed, freely moving counterions \cite{Hessinger2000, Medebach2005}.
Although the various effective charges, probed by these different experiments, are conceptually
different from each other, the ratio of their numerical values seems to be correlated \cite{Medebach2005, Medebach2007, Shapran2005}.  
The chemical and experimental boundary conditions for charged sphere suspensions can be varied over a wide range \cite{Yethiraj2007}, allowing
for large variations in the colloidal charge numbers.
In a self-consistent parameter-free approach, the colloidal bare and effective charges in an aqueous solvent
should be predicted based on the chemical equilibrium conditions of dissociated surface ionic groups and bulk ions \cite{Doi2013}.

Nonlinear screening theories \cite{Lowen1993, Lowen1992},
computer simulations of the primitive model
\cite{Linse1999, Lobaskin1999, Allahyarov1998, Lowen1998, Allahyarov1998b}
and liquid integral equation theory of strongly coupled Coulomb systems
\cite{Khan1987, Belloni1986, Heinen2013}
are routinely used to treat the ionic correlations, but the second aspect of bare charge variability
has often been ignored in these approaches.

Monte Carlo \cite{Pellenq1997, Khan2005, Moreira2002, Labbez2009, Madurga2011, Barr2011}
or Molecular Dynamics \cite{Messina2001, Calero2010} computer simulations with an explicit account for 
charged surface groups are computationally very expensive, especially when the size- and charge disparity between macroions and microions is large.
This renders the development of computationally more efficient methods desirable \cite{Teixeira2010}. 
For a recent review on surface charge regulation in biomolecular solutions, we refer to \refname{}~\cite{Lund2013}.

Behrens, Borkovec and Grier have solved the problem of charge regulation of two electrolyte-immersed surfaces,
with Poisson-Boltzmann microion distributions \cite{Behrens1999a, Behrens1999b, Behrens1999c, Behrens2001a} and, recently,
the conductivity of charged, electrolyte-filled fluidic nanochannels has been investigated in a comparable
mean-field-level study \cite{Fleharty2014}.
Coupled surface and bulk chemistry in colloidal suspensions has also been considered in a mean-field-like approach
\cite{Carrique2001, Carrique2003, Ruiz-Reina2008, Carrique2009, Carrique2010}
which takes account of macroion correlations only within (revised versions of)
the minimalistic cell model \cite{Alexander1984}. This was used to predict
electrokinetic properties of aqueous suspensions. An account of water self-dissociation
and carbon dioxide based contaminations yielded an improved
agreement with experimental data in these studies on the mean-field level.

In this paper we tackle both problems -- the non-mean-field correlations in ionic colloidal suspensions \textit{and}
the chemical regulation of the colloidal charge -- simultaneously, in a self-consistent semi-analytical approach based on liquid integral equations.
Thereby, ionic correlations beyond the linear screening theory level are incorporated in a good approximation.
At the same time, the liquid integral equation solution provides a coupling between the chemical association-dissociation balances
of acidic groups on the colloidal sphere's surfaces, and the bulk concentrations of all ionic species.
\begin{figure}
 \includegraphics[width=.6\columnwidth,angle=-90]{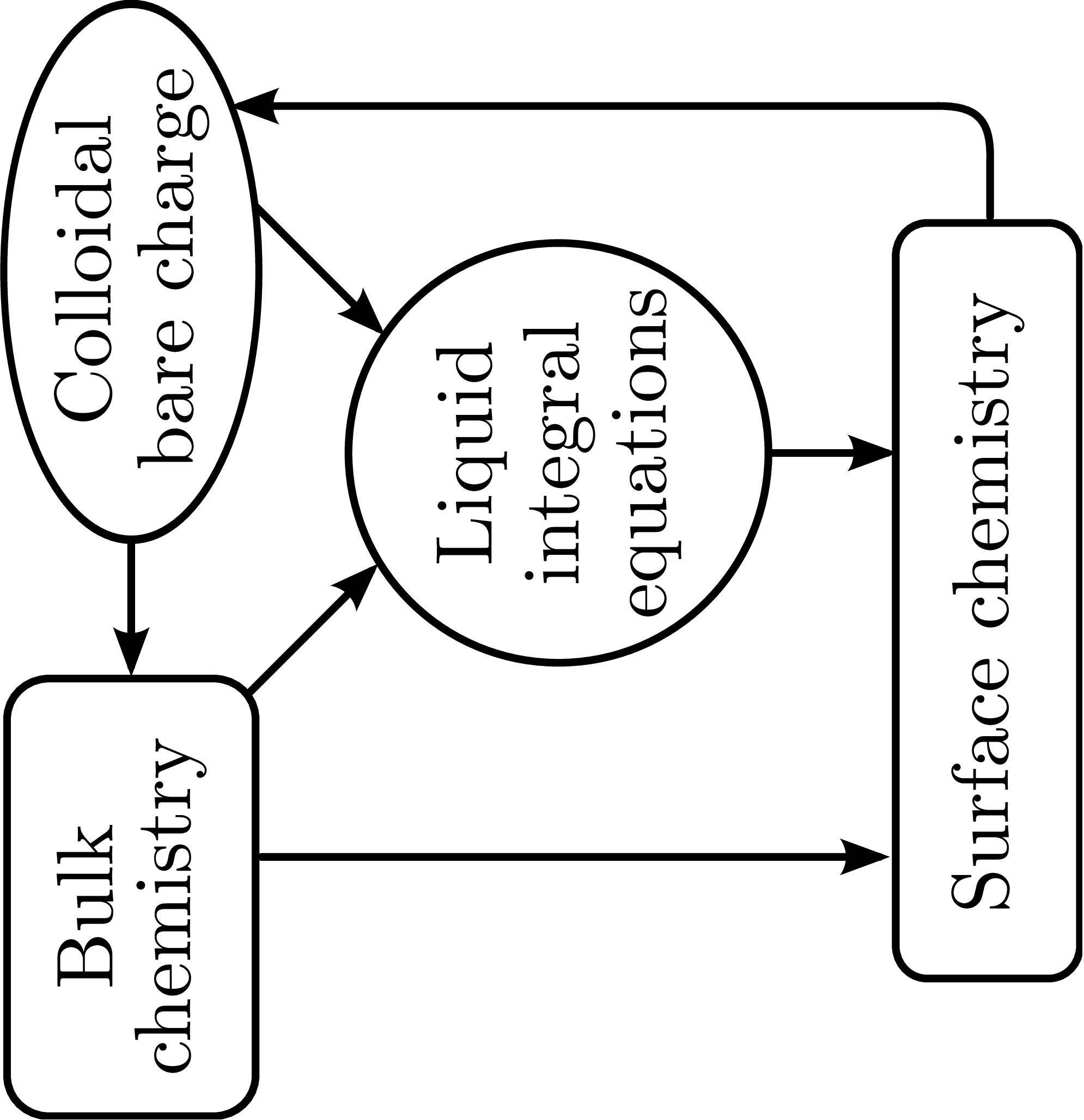} 
 \vspace{1em}
 \caption{Schematic diagram of the coupled physicochemical problem.
 Chemical association-dissociation balances in bulk suspension and at the colloidal particle's surfaces
 couple to the electrostatic and statistical-mechanical problems of variable colloidal bare charge and particle pair-correlations
 described in the liquid integral equation approach. Relations between the four subproblems, that are taken account of in the
 present work, are indicated by arrows. A closed graph of subproblems is obtained that can be self-consistently solved.  
 }
 \label{fig:schematic}
\vspace{1em}
\end{figure}
The key idea, illustrated schematically in \figurename~\ref{fig:schematic}, is that liquid integral equations predict the 
excess chemical potentials of all ionic species, which then enter into the chemical association-dissociation balance of colloidal
acidic surface groups. The degree of surface-group
dissociation is directly proportional to the colloidal bare charge, which, in turn, influences the
overall (bulk) ionic composition and the pair correlations among all ion species in the liquid integral equation system.
The so-obtained implicit set of physicochemical equations is numerically self-consistently solved, yielding results that
include colloidal bare and effective charges, and the suspension's $pH$-value.

Note that the major difficulty in tackling the coupled equation set
lies in the numerical solution of the involved liquid integral equations.
When all ion species are treated on equal footing in the so-called primitive model, as done in the present work,
very large asymmetries between the (effective) hard-core diameters and charge numbers of macro- and microions must be resolved.
These asymmetries pose a formidable challenge for the numerical stability
and efficiency of solution methods for liquid integral equations. Solving the equations 
that occur in the present study within reasonable program execution times has been rendered possible only recently,
with the advent of a numerical solution method by part of the present authors \cite{Heinen2013}. This method is based on
earlier work by different groups \cite{Ng1974, Talman1978, Rossky1980, Hamilton2000, Hamilton_website}, the key ideas of
which have been generalized and combined in a versatile way.

This paper is organized as follows:
In \sectionname~\ref{sec:Theory}, we explicate our theoretical scheme, including
association-dissociation balances between all relevant reactive species in 
\sectionname~\ref{sec:sub:asso-disso}, constraints on the number concentrations
in \sectionname~\ref{sec:sub:constraints}, ion pair-correlations
and activities in \sectionname~\ref{sec:sub:HNC} and \appendixname~\ref{AppendixA},
the effective charge number of colloidal spheres in \sectionname~\ref{sec:sub:Zeff}, 
and the self-consistent solution of the coupled physicochemical equation set in
\sectionname~\ref{sec:sub:sel-cons} and \appendixname~\ref{AppendixB}.
Results predicted by our theoretical scheme are presented in \sectionname~\ref{sec:Results},
beginning with a discussion of macro- and microion pair-correlation functions in \sectionname~\ref{sec:sub:Paircorr},
and macroion bare and effective charges as well as the suspension's $pH$-value in \sectionname~\ref{sec:sub:Zeff_Results}.
In the final two sections \ref{sec:Outlook} and \ref{sec:Conclusions}, we mention possible future
continuations and extensions of the present work, and give our concluding remarks.

\section{Theoretical scheme}\label{sec:Theory}

In the following, we investigate aqueous suspensions of monodisperse colloidal spheres in thermodynamic equilibrium.
Each colloidal sphere carries a mean (time-averaged)
electric charge of magnitude $Ze$, where $Z$ is the colloidal bare charge number, and $e$ denotes the proton
elementary charge. In the model description applied here, colloidal spheres acquire their electric charge solely
by the dissociation of acidic surface groups that are covalently bound to the sphere surfaces.
We have limited our studies to two types of colloidal spheres, with either strongly or weakly acidic surface groups.
The first type represents spheres that are covered with strongly acidic sulfonate \mbox{(R-O-SO$_3$H)} surface groups. Such particles
have been synthesized and used in various experimental studies of phase behavior, equilibrium and non-equilibrium properties
\cite{Sood1991, Medebach2005, Westermeier2012}.
The second type represents spheres covered with weakly acidic silanol \mbox{(SiOH)} surface groups, such as the experimentally frequently
used colloidal silica particles \cite{PhilipseVrij1988, Yoshida1999, Gapinski2009, Wette2010, Heinen2011}.
Weakly acidic surface groups allow for a considerable variation of the colloidal charge by altering the suspension parameters.
Both kinds of surface groups are monovalent acids. As a consequence, we have $0 \geq Z \geq -N$.

We denote the number concentration of particles of species i by [i], and all number concentrations in this paper
are given in units of \mbox{M $=$ 1 mol$/$liter}. The (bulk) number concentration [i] is defined as the total number of particles of species i,
divided by the total system volume.
In the primitive model (PM) description applied here, spherical colloidal macroions as well as monovalently charged microions
are approximated as non-overlapping hard spheres with pairwise additive hard-core diameters.
The methods presented in this paper could in principle be applied to suspensions including multivalent low-molecular-weight
microions, where effective charge inversion of colloidal spheres has been observed 
\cite{Kuhn1999, Shklovskii1999, Nguyen2000, Nguyen2000PRL, Martin-Molina2003, Besteman2004, Quesada-Perez2005, Zhang2008, Calero2009, Roosen-Runge2013}.
However, for the sake of simplicity we limit ourselves here to suspensions with monovalent microions. 
Throughout our analysis, we assume the approximate microion effective sphere diameters
\begin{eqnarray}
\sigma_{\text{H}_3\text{O}^+} = \sigma_{\text{OH}^-} &=& 0.9~\text{nm}\qquad\text{and}\\~\nonumber\\
\sigma_{\text{HCO}_3^-} &=& 1.1~\text{nm},
\end{eqnarray}
reminiscent of ions dressed with one hydration layer of H$_2$O molecules. All results presented here are for colloidal spheres with hard core diameter
\begin{equation}
\sigma_{\text{Col}} = 100~\text{nm}.
 \end{equation}

\subsection{Association-dissociation balances}\label{sec:sub:asso-disso}

\begin{figure}
 \includegraphics[width=.5\columnwidth,angle=-90]{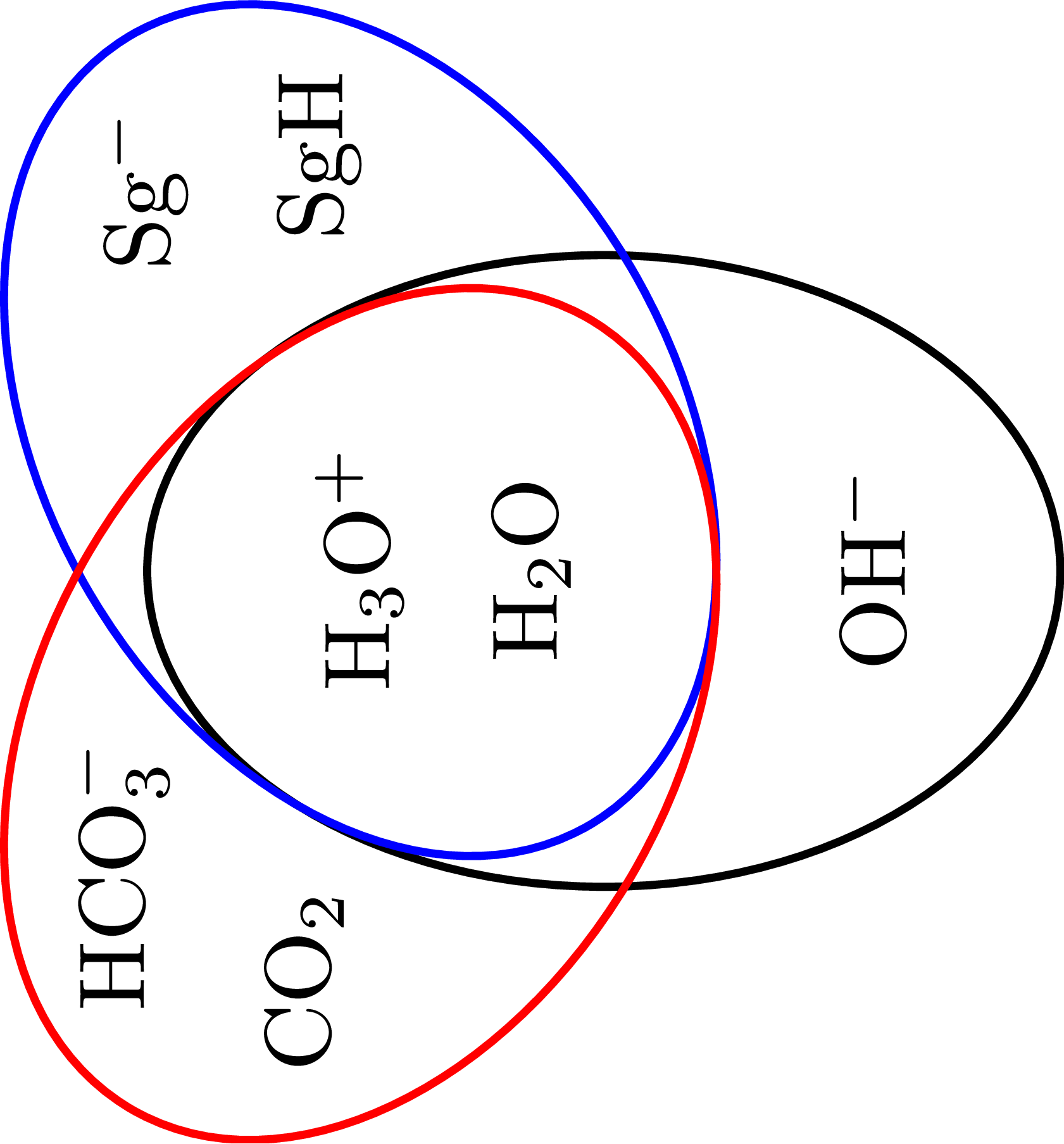} 
 \caption{(Color online) All relevant chemically reactive species in suspension. Here, SgH stands for one acidic colloidal surface group
 (sulfonate or silanol in the cases studied here), and Sg$^-$ is one deprotonated surface group.
 Reactants of each of the three association-dissociation balances in \expressionsname~\eqref{eq:CO2_reaction}-\eqref{eq:surfgroup_reaction}
 are surrounded by one individual ellipse.
 }
 \label{fig:coupled_reaction_species}
\vspace{1em}
\end{figure}

In \figurename~\ref{fig:coupled_reaction_species}, the chemical formulas of the seven reactive species of interest are given, including
water (H$_2$O), carbon dioxide (CO$_2$), bicarbonate (HCO$_3^-$), hydronium (H$3$O$^+$), hydroxide (OH$^-$), and colloidal surface groups
with (SgH) or without (Sg$^-$) an attached proton. For the systems studied in the following, SgH either stands for one sulfonate (R-O-SO$_3$H)
or one silanol (SiOH) group. The species in \figurename~\ref{fig:coupled_reaction_species} are grouped by three ellipses, each surrounding the
reactants of one of the three fundamental association-dissociation balances of the system, which are 
\begin{eqnarray}
~~~\text{CO}_2 + 2\text{H}_2\text{O} &\stackrel{pK_a^{\text{CO}_2}}{\leftrightharpoons}& \text{H}_3\text{O}^+ + \text{HCO}_3^-, \label{eq:CO2_reaction}\\~\nonumber\\
2 \text{H}_2\text{O} &\stackrel{pK}{\leftrightharpoons}& \text{H}_3\text{O}^+ + \text{OH}^-,~\text{and} \label{eq:water_selfdissociation}\\~\nonumber\\
\text{SgH} + \text{H}_2\text{O} &\stackrel{pK_a^{\text{SgH}}}{\leftrightharpoons}& \text{H}_3\text{O}^+ + \text{Sg}^-.\label{eq:surfgroup_reaction}\\~\nonumber
\end{eqnarray}
Here, $pK_a^{\text{SgH}}$ and $pK_a^{\text{CO}_2}$ are the acid dissociation constants of the surface groups and of carbon dioxide, respectively, and $pK$ is
the water self-dissociation constant.
Note that \expressionname~\eqref{eq:CO2_reaction} is short-hand notation for
the combined two reactions CO$_2$ + H$_2$O $\leftrightharpoons$ H$_2$CO$_3$ and H$_2$CO$_3$ + H$_2$O $\leftrightharpoons$ HCO$_3^-$ + H$_3$O$^+$, proceeding via the
intermediate species carbonic acid (H$_2$CO$_3$). Carbonic acid molecules are electrically neutral, and therefore do not influence the PM ion pair-correlation functions,
discussed further down in subsection~\ref{sec:sub:HNC}. Also, H$_2$CO$_3$ molecules do not directly participate
in either of the two reactions in \expressionsname~\eqref{eq:water_selfdissociation} and \eqref{eq:surfgroup_reaction}. It is thus unnecessary
to include carbonic acid molecules explicitly into our description.

The \mbox{CO$_2$} dissociation constant $pK_a^{\text{CO$_2$}}$ quantifies the
equilibrium thermodynamic activity ratio \cite{HillBook, MooreBook}
\begin{equation}\label{eq:CO2_balance_pKa}
\frac{\displaystyle {a_{\text{HCO$_3^-$}} {a_{\text{H$_3$O$^+$}}}} }{\displaystyle{a_{\text{CO$_2$}}}} = K_a^{\text{CO$_2$}} = 10^{-pK_a^{\text{CO$_2$}}} \text{M}
\end{equation}
for the reaction in \expressionname~\eqref{eq:CO2_reaction}.
In \expressionname~\eqref{eq:CO2_balance_pKa} and further down this text, the thermodynamic activity, $a_{\text{i}}$, of species i
is defined according to the the convention
\begin{equation}\label{eq:activity_definition}
\beta \mu_{\text{i}} = \ln\left(a_{\text{i}} \Lambda_{\text{i}}^3 \right) =
\ln\left([{\text{i}}] \gamma_{\text{i}} \Lambda_{\text{i}}^3 \right) =
\beta \mu_{\text{i}}^{\text{exc}} + \beta \mu_{\text{i}}^{\text{id}},    
\end{equation}
where $\beta = 1/(k_B T)$ with Boltzmann constant $k_B$ and absolute Temperature $T$, and where $\mu_{\text{i}}$ denotes the chemical potential of species i.
The latter can be written as the sum of the excess chemical potential $\mu_{\text{i}}^{\text{exc}} = \ln(\gamma_{\text{i}})/\beta$
and the ideal chemical potential $\mu_{\text{i}}^{\text{id}} = \ln([{\text{i}}]\Lambda_{\text{i}}^3)/\beta$.
In \expressionname~\eqref{eq:activity_definition}, $\gamma_{\text{i}} = a_{\text{i}} / [{\text{i}}]$ is the activity coefficient, and
$\Lambda_i$ is the thermal de Broglie wavelength, which is of no relevance in the following.
Employing the conventions in \expressionname~\eqref{eq:activity_definition} implies that the reference state of substance i is an
ideal gas at number density $[{\text{i}}]$, with chemical potential $\mu_{\text{i}}^{\text{id}}$.

In the following, we approximate $\gamma_{\text{i}} = 1$ for all electrically neutral species i.
Then, \expressionname~\eqref{eq:CO2_balance_pKa} can be re-written as
\begin{equation}\label{eq:CO2_balance_gammas}
\frac{\displaystyle{\text{[HCO$_3^-$]} \times \text{[H$_3$O$^+$]}}}{\displaystyle{\text{[CO$_2$]}}} =
\frac{\displaystyle{K_a^{\text{CO$_2$}}}}{\displaystyle{\gamma_{\text{HCO$_3^-$}} \gamma_{\text{H$_3$O$^+$}}}},
\end{equation}
and the analogous equation
\begin{equation}\label{eq:water_balance_gammas}
\frac{\displaystyle{\text{[OH$^-$]} \times \text{[H$_3$O$^+$]}}}{\displaystyle{\text{[H$_2$O]}}} =
\frac{\displaystyle{K}}{\displaystyle{\gamma_{\text{OH$^-$}} \gamma_{\text{H$_3$O$^+$}}}}
\end{equation}
quantifies the equilibrium state of the water self dissociation
reaction in \expressionname~\eqref{eq:water_selfdissociation}, with $K = 10^{-pK} \text{M}$.

The equilibrium state of the acidic surface group dissociation reaction in \expressionname~\eqref{eq:surfgroup_reaction}
is characterized by
\begin{equation}\label{eq:surfgroup_balance_pKa}
\frac{\displaystyle {a_{\text{H$_3$O$^+$}} {a_{\text{Sg$^-$}}}} }{\displaystyle{a_{\text{SgH}}}} = K_a^{\text{SgH}} = 10^{-pK_a^{\text{SgH}}} \text{M}.
\end{equation}
Realizing that $pH = -\log_{10}(a_{\text{H$_3$O$^+$}})$, $\text{[Sg$^-$]} = |Z| \text{[Col]}$ and $\text{[SgH]} = (N-|Z|) \text{[Col]}$,
\expressionname~\eqref{eq:surfgroup_balance_pKa} can be converted into the Henderson-Hasselbalch equation
\begin{equation}\label{eq:Henderson_Hasselbalch}
\log_{10} \left(\frac{\displaystyle{|Z|/N}}{\displaystyle{1-|Z|/N}} \right) = pH -pK_a^{\text{SgH}} - \log_{10}(\gamma_{\text{Sg$^-$}}),
\end{equation}
quantifying the chemical regulation of $Z$.

Links between the three \expressionsname~\eqref{eq:CO2_balance_gammas},~\eqref{eq:water_balance_gammas} and \eqref{eq:Henderson_Hasselbalch}
are provided by the hydronium ion concentration, $\text{[H$_3$O$^+$]}$, and also by the four activity coefficients
$\gamma_{\text{H$_3$O$^+$}}$, $\gamma_{\text{OH$^-$}}$, $\gamma_{\text{HCO$_3^-$}}$, and $\gamma_{\text{Sg$^-$}}$, each of which
depends on the charge and concentration of all ionic species in suspension. In the self-consistent PM solution scheme used here, the intricate relations
between the $\gamma_i$'s, [i]'s and $Z$ are resolved within the HNC approximation (\textit{c.f.}, \sectionname~\ref{sec:sub:HNC}).

All results presented in the following have been obtained using the (acid) dissociation constants
\begin{eqnarray}
pK &=& 15.74,\rule[-3mm]{0mm}{6mm}\\
pK_a^{\text{CO}_2} &=& 6.5,\rule[-3mm]{0mm}{6mm}\\
pK_a^{\text{SiOH}} &=& 4.0,\rule[-3mm]{0mm}{6mm}\qquad\text{and}\label{eq:silanol_pKa}\\
pK_a^{\text{R-O-SO}_3\text{H}} &=& 1.5,\rule[-3mm]{0mm}{6mm}\label{eq:sulfate_pKa}
\end{eqnarray}
for water, carbon dioxide, silanol, and sulfonate dissociation, respectively.
Values in \expressionsname~\eqref{eq:silanol_pKa} and \eqref{eq:sulfate_pKa} were chosen as typical representative cases
of a weak and a strong acid.

\subsection{Concentration constraints}\label{sec:sub:constraints}

Without further constraints, the three \expressionsname~\eqref{eq:CO2_balance_gammas},~\eqref{eq:water_balance_gammas} and \eqref{eq:Henderson_Hasselbalch},
containing five number concentrations, four different activity coefficients, and the unknown charge number $Z$, do not possess an unambiguous solution.
In the following, we construct a closed set of equations with a unique solution by identifying the relevant
concentration constraints, and by providing PM-HNC expressions for the activity coefficients.
We begin by identifying the known and unknown quantities, listed in \tablename~\ref{tab:known_and_unknown}.

\begin{table}
\caption{The known input parameters of our suspension model, and the basic quantities that need to be determined for an unambiguous solution.
[Col] is the number concentration of colloidal spheres, each carrying $N$ surface groups, $|Z|$ of which are dissociated in
equilibrium. Further explanations are given in subsection~\ref{sec:sub:constraints}}\label{tab:known_and_unknown}
\vspace{1em}
\centering
{
\begin{tabular}{@{\extracolsep{\fill}}lll}
\hline
\textbf{Known:}\qquad\qquad\qquad\qquad\qquad\qquad&\textbf{Unknown:}\\ 
\hline\hline
$N$,                                                   & $Z$,                               & $\gamma_{\text{Sg$^-$}}$,\rule[-3mm]{0mm}{6mm}\\
\text{[Col]},                                          & \text{[OH}$^-$\text{]},            & $\gamma_{\text{OH$^-$}}$,\rule[-3mm]{0mm}{6mm}\\
\text{[H}$_2$\text{O]}$ \equiv 54.2$ M,                & \text{[H}$_3$\text{O}$^+$\text{]},\qquad\qquad\qquad & $\gamma_{\text{H$_3$O$^+$}}$,\rule[-3mm]{0mm}{6mm}\\
\text{[CO}$_2$\text{]}$ \equiv 1.52 \times 10^{-5}$ M, & \text{[HCO}$_3^-$\text{]},         & $\gamma_{\text{HCO$_3^-$}}$, \rule[-3mm]{0mm}{6mm}\\
\hline
\hline
\end{tabular}
}
\label{tab:simparam}
\end{table}

In the left column of \tablename~\ref{tab:known_and_unknown}, the relevant known input parameters are listed, beginning with the number, $N$, of dissociable
surface groups per colloidal sphere. This quantity is assumed to be known since, in typical experiments, it can be accurately determined by titration
\cite{Gisler1994, Yamanaka1997, Hessinger2000}.
Likewise, the number concentration of colloidal spheres, [Col], is assumed to be known since it is an experimentally rather
well-controlled quantity. It can either be measured directly \cite{Luck1963, Wette2002, Wette2010}, or it 
can be calculated, \textit{e.g.}, on basis of a colloidal form-factor measurement, the colloidal sphere mass density,
and the colloidal mass fraction \cite{Westermeier2012}. In presenting our results for different values of [Col] in \sectionname~\ref{sec:Results},
we use the colloidal volume fraction
\begin{equation}\label{eq:volfrac_definition}
\phi = \frac{\pi}{6} \sigma_{\text{Col}}^3 \text{[Col]}, 
\end{equation}
as a control parameter, since $\phi$ is more intuitively interpreted than the quantity [Col].
In \expressionname~\eqref{eq:volfrac_definition}, $\phi$ is the fraction of the total suspension volume that is occupied by colloidal spheres.

Since water molecules are the overwhelming majority species, it is a good approximation to assume a constant [H$_2$O] $ = 54.2$ M, which corresponds
to the number concentration of pure water. This concentration is many orders of magnitude higher than that of any other species in the self-consistent
solutions reported in \sectionname~\ref{sec:Results}.

As regards carbon dioxide, we assume a concentration of
[CO$_2$]$ = 1.52 \times 10^{-5}$ M, which corresponds to CO$_2$-saturated, salt-free water under an atmosphere with a CO$_2$ partial pressure
of $3.9 \times 10^{-4}$ atm. \cite{Weiss1974, Millero1995}. Note here again that $\gamma_{\text{CO}_2}$ is equal to one in our approximate description.
It is therefore consistent to prescribe the number concentration of CO$_2$.

In addition to fixing [H$_2$O] and [CO$_2$], a constraint arises from requiring global electroneutrality
of the suspension, which can be written as
\begin{equation}\label{eq:global_en}
Z \text{[Col]} + \text{[H}_3\text{O}^+\text{]} - \text{[OH}^-\text{]} - \text{[HCO}_3^-\text{]}  = 0. 
\end{equation}
The global electroneutrality constraint in \expressionname~\eqref{eq:global_en}, combined with \expressionsname~\eqref{eq:CO2_balance_gammas} and \eqref{eq:water_balance_gammas},
gives the quadratic equation
\begin{equation}\label{eq:quadratic_eqn_for_HCO3-}
\text{[H}_3\text{O}^+\text{]}^2 - Z \text{[Col]} \text{[H}_3\text{O}^+\text{]} =
\frac{\displaystyle{K_a^{\text{CO}_2} \text{[CO}_2\text{]}}}{\displaystyle{\gamma_{\text{HCO$_3^-$}} \gamma_{\text{H$_3$O$^+$}}}} +
\frac{\displaystyle{K \text{[H}_2\text{O]}}}{\displaystyle{\gamma_{\text{OH$^-$}} \gamma_{\text{H$_3$O$^+$}}}},  
\end{equation}
with a unique physical (positive) solution for $\text{[H}_3\text{O}^+\text{]}$.

At this point we have collected the four
\expressionsname~\eqref{eq:CO2_balance_gammas},~\eqref{eq:water_balance_gammas}, \eqref{eq:Henderson_Hasselbalch} and \eqref{eq:quadratic_eqn_for_HCO3-}.
In combination with the HNC scheme solution, from which the ion activity coefficients are obtained,
these equations are sufficient to determine all eight unknowns listed in the right column of \tablename~\ref{tab:known_and_unknown}.

\subsection{HNC scheme}\label{sec:sub:HNC}

We employ the liquid integral equation formalism to compute the pair-correlations among all ionic species in suspension,
based on the multicomponent Ornstein-Zernike (OZ) equations \cite{Hansen_McDonald1986}
\begin{equation}\label{eq:O-Z}
h_{i,j}(r) = c_{i,j}(r) + \sum\limits_k [k] \int d^3 {\boldsymbol{r}}' c_{i,k}(r') h_{k,j}(r-r'),
\end{equation}
which are valid for a homogeneous and isotropic, three-dimensional fluid mixture.
In \expressionname~\eqref{eq:O-Z}, the $c_{i,j}(r)$ and $h_{i,j}(r) = g_{i,j}(r) -1$ are the partial direct and total correlation functions, respectively,
between ions of species $i$ and $j$.

Here, we solve the coupled OZ equations for a system of \textit{five} ionic species:
Number one to four are the species H$_3$O$^+$, HCO$_3^-$, OH$^-$ and Col, the latter denoting entire colloidal spheres that carry a charge of $Ze$ each.
The fifth ionic species is identified by the lower index 'dilCol' in the following, and represents an ultradilute fluid of colloidal spheres
with diameter $\sigma_{\text{Col}}$ and with a charge of $(Z-1)e$.
Species dilCol is introduced merely as a bookkeeping device, necessary for the determination of the surface group excess chemical potential
$\mu_{\text{Sg$^-$}}^{\text{exc}}$, as explicated in \appendixname~\ref{AppendixA}. The number concentration [dilCol] is selected several orders of magnitude
smaller than [Col]. Hence, species dilCol exerts a negligible influence on the mutual pair-correlation functions between the four
species H$_3$O$^+$, HCO$_3^-$, OH$^-$ and Col.

To obtain a closed set of integral equations, \expressionsname~\eqref{eq:O-Z} are combined with the approximate HNC closure relation \cite{Morita1958, Hansen_McDonald1986}
\begin{equation}\label{eq:HNC}
g_{i,j}(r) = \exp\left\lbrace - u_{i,j}(r) + h_{i,j}(r) - c_{i,j}(r) \right\rbrace, 
\end{equation}
in which the $u_{i,j}(r)$ are the dimensionless pair-potentials of direct interaction between ions,
\begin{equation}\label{eq:pair_pot}
u_{i,j}(r) = \left\lbrace
   \begin{array}{ll}
   \infty\,& ~~\text{for}~r < \sigma_{i,j},\\~\\
   \frac{\displaystyle{L_B Z_i Z_j}\rule{0em}{1em}}{\displaystyle{r}\rule{0em}{.9em}}\,& ~~\text{for}~r > \sigma_{i,j},\\
   \end{array}
 \right. \, \\
\end{equation}
invoking the solvent-characteristic Bjerrum length $L_B = e^2/(\epsilon k_B T)$ in Gaussian units and
the pairwise additive hard core diameters $\sigma_{i,j} = (\sigma_i + \sigma_j)/2$.
In all calculations with results presented here, we have used $L_B = 0.701$ nm, corresponding to water at room temperature.
Assuming pair potentials of the kind of \expressionname~\eqref{eq:pair_pot} for
the microion and macroion species, amounts to an approximate treatment of the ion pair-interactions within the PM.

The PM description neglects short-ranged van der Waals attraction,
as well as changes in water polarizability which can play a role a high surface potential \cite{Hatlo2012}.
Furthermore, it is assumed that the charge of a colloidal sphere is homogeneously smeared out over the sphere surface.
Our model thus neglects all effects arising from charge patchiness \cite{Messina2001},
a topic that has recently received much interest in studies based on the nonlinear and anisotropic Poisson-Boltzmann equation \cite{Boon2010, Boon2011, deGraaf2012}.
Surface charge patchiness could in principle be included into our description, if the OZ \expressionsname~\eqref{eq:O-Z}
were replaced by a reference interaction site model \cite{Andersen1970, Hansen_McDonald1986, Schweizer1997, Harnau2002} description,
or by anisotropic OZ equations \cite{Brandt2010}.
However, the strong charge- and diameter asymmetry between macroions and microions
renders already the solution of \expressionsname~\eqref{eq:O-Z}-\eqref{eq:pair_pot} into a tedious task \cite{Leger2005, Heinen2013}.

We solve \expressionsname~\eqref{eq:O-Z}-\eqref{eq:pair_pot} by means of our
recently developed method \cite{Heinen2013}, which is specially well-suited for application to
highly asymmetric electrolytes, in an arbitrary number of spatial dimensions.
For details of the solution method, which relies on a generalized version of Ng's fixed point iteration scheme \cite{Ng1974}
and a Fourier-Bessel transform on computational grids with logarithmic
spacing \cite{Talman1978, Hamilton2000, Hamilton_website}, we refer to our comprehensive description in \refname~\cite{Heinen2013}. Note here that
essentially the same numerical method has been used already in the year 1980 by Rossky and
Friedman \cite{Rossky1980}. Our algorithm constitutes an optimization and generalization of this earlier
work, and the first application of the method to highly asymmetric electrolytes. 

Once that \expressionsname~\eqref{eq:O-Z}-\eqref{eq:pair_pot} have been solved for a given set of $[k]$'s and a given $Z$, the correlation functions
are used as input for computing the thermodynamic activity coefficients of all ionic species by means of the Hansen-Vieillefosse-Belloni equation
\cite{Hansen1976, Hansen1977, Belloni1985, Hansen_McDonald1986, Hopkins2006, Gutierrez-Valladares2011}
\begin{eqnarray}
\ln(\gamma_i) = \beta\mu_i^{\text{exc}} &=& \sum\limits_j [j] \int d^3 {\boldsymbol{r}} ~
\frac{1}{2} h_{ij}(r)\left[h_{ij}(r) - c_{ij}(r)\right]\nonumber\\
&-& \sum\limits_j [j] \int d^3 {\boldsymbol{r}}~  \left[c_{ij}(r) +u_{ij}(r)\right].\label{eq:Hansen-Vieillefosse-Belloni}
\end{eqnarray}
The surface group excess chemical potential, $\ln(\gamma_{\text{Sg}^-})$,
which is the essential quantity in colloidal surface charge regulation described by \expressionname~\eqref{eq:Henderson_Hasselbalch},
is obtained within the PM as the right-hand-side of \expressionname~\eqref{eq:surfgroup_exc_chempot}. It is taken as the sum of 
the colloidal sphere Coulomb self-energy change, caused by the dissociation of one surface group,
plus the difference between the excess chemical potentials of colloidal spheres with charges $(Z-1)e$ and $Ze$.

A brief discussion is in place here, regarding the accuracy of \expressionname~\eqref{eq:Hansen-Vieillefosse-Belloni}, which is the HNC approximation of an exact expression
that has been derived by Kjellander and Sarman \cite{Kjellander1989} and Lee \cite{Lee1992} (see also \refname{}~\cite{Sarkisov2001}). 
In \refsname{}~\cite{Lee1992} and \cite{Bomont2004} it has been shown and discussed that \expressionname~\eqref{eq:Hansen-Vieillefosse-Belloni}
generally provides a very poor approximation for the excess chemical potential of particles with a hard core. Since we are indeed concerned with particles 
that exhibit hard-core plus Coulomb interactions, the applicability of \expressionname~\eqref{eq:Hansen-Vieillefosse-Belloni} may therefore be questioned.
However, our method for calculating the salient surface group excess chemical potential is based on the \emph{difference}
$\mu_{\text{dilCol}}^{\text{exc}} - \mu_{\text{Col}}^{\text{exc}}$ between excess chemical potentials of colloidal spheres that differ
in their electric charges, but not in their hard core diameters.
As we have checked, the inaccurate hard-core contributions (\textit{i.e.}, the contributions to the integrals
in \expressionname~\eqref{eq:Hansen-Vieillefosse-Belloni} for $0 < |\boldsymbol{r}| < \sigma_{\text{Col, HCO$_3^-$}}$) are practically identical
for both species Col and dilCol, and therefore cancel out nearly perfectly when the difference is taken.
The remaining non-overlap parts of the integrals in \expressionname~\eqref{eq:Hansen-Vieillefosse-Belloni} are quite accurate due to   
the very rapid decay of the (neglected) bridge function at non-overlap distances of particles with Coulomb interactions.

In addition to the surface group excess chemical potential, the hydronium ion excess chemical potential
$\ln(\gamma_{\text{H$_3$O$^+$}})$ enters into the charge regulation \expressionname~\eqref{eq:Henderson_Hasselbalch},
via $pH = -\log_{10}(\gamma_{\text{H$_3$O$^+$}} [\text{H$_3$O$^+$}])$. In computing $\ln(\gamma_{\text{H$_3$O$^+$}})$,
the inaccurate hard-core contributions to the integrals in \expressionname~\eqref{eq:Hansen-Vieillefosse-Belloni} play no
significant role either, due to two reasons: First, the number concentration $\text{[Col]}$ is orders of magnitude smaller than
$[\text{H$_3$O$^+$}]$ in all examples studied here, such that the summands with \mbox{j $=$ Col} play no important role
for \mbox{i $=$ H$_3$O$^+$}. Second, as we have numerically tested, the remaining relevant microion-microion contributions to the sums
in \expressionname~\eqref{eq:Hansen-Vieillefosse-Belloni} are totally dominated by the electrostatic (non-overlap)
parts of the integrals, due to the strong electrostatic interactions amongst microions.

In a future extension of the present work, the HNC closure may be replaced by a thermodynamically partially consistent
closure relation. Here, a  specially suitable candidate is the closure that has been proposed by Bomont and Bretonnet
\cite{Bomont2003a}, and that has been supplemented by an expression for the excess chemical potential \cite{Bomont2003b, Bomont2004},
similar in form to \expressionname~\eqref{eq:Hansen-Vieillefosse-Belloni}, but significantly less suffering from an inaccurate hard-core contribution.
Bomont and Bretonnet's closure is especially well suited for application to a restricted PM of electrolytes containing
microions only, or for electrolytes containing rather small polyions like, \textit{e.g.}, charged globular proteins
\cite{Heinen2012}. Note, however, that the application of a thermodynamically self-consistent closure to a PM with strong
charge- and size-asymmetries is somewhat hampered by the fact that
the number of correlation functions raises more quickly than the number of consistency criteria when the number of species is increased \cite{Belloni1988}.
Therefore, keeping in mind the slight inaccuracy of \expressionname~\eqref{eq:Hansen-Vieillefosse-Belloni},
we resort to the simpler HNC scheme in the present work.

\subsection{Colloidal effective charge}\label{sec:sub:Zeff}

In the analysis of experiment results,
and in the construction of theoretical schemes for colloidal dynamics,
one is often interested in a mesoscopic description of reduced complexity,
where the microion's degrees of freedom have been integrated out. 
In such a one-component macroion fluid (OMF) description, the colloidal spheres
remain as the only species whose correlations are explicitly resolved, and the
hard-sphere Coulomb pair-potential among macroions,
$u_{\text{Col, Col}}(r)$,
must be replaced by an effective, state-dependent macroion pair potential
$u_{\text{Col, Col}}^{\text{eff}}(r)$
that takes implicit account of the presence of microions.

Having solved the coupled PM-HNC \expressionsname~\eqref{eq:O-Z}-\eqref{eq:pair_pot} for all ionic species,
an effective macroion pair potential can be extracted via an inversion of the HNC relation \cite{Fushiki1988, Heinen2013}.
In a very similar way, HNC inversion has been used to extract effective macroion potentials from digital
video microscopy data \cite{Behrens2001b}.
The effective macroion potential from HNC inversion can be mapped to the electrostatic repulsive part,
\begin{equation}\label{eq:u_DLVO}
u^{\text{DLVO}}_{\text{Col, Col}}(r) =
L_B \left( \frac{\displaystyle{Z_{\text{eff}} e^{\displaystyle{\kappa a_{\text{Col}}} }}}{\displaystyle{1 + \kappa a_{\text{Col}}}} \right)^2
\dfrac{{e^{-\kappa r}}}{r},\qquad r > \sigma_{\text{Col, Col}} 
\end{equation}
of the Derjaguin-Landau-Verwey-Overbeek (DLVO) pair potential between two finite-sized macroions in an electrolyte with microion correlations
treated in Debye-H\"{u}ckel approximation \cite{Verwey_Overbeek1948}. For the chemical composition of suspensions studied here,
the square of the inverse exponential screening length $\kappa$ in \expressionname~\eqref{eq:u_DLVO} is given by 
\begin{equation}\label{eq:kappa}
\kappa^2 = 4 \pi L_B\;\! \left( {\text{[Col]}} |Z_{\text{eff}}| + 2 {\text{[HCO}_3^-\text{]}} + 2 {\text{[OH}^-\text{]}} \right).
\end{equation}
In case of a dilute suspension of weakly charged macroions with
$|L_B Z / \sigma_{\text{Col, Col}}| \ll 1$, the potential in \expressionname~\eqref{eq:u_DLVO} accurately represents the 
effective macroion pair potential with $Z_{\text{eff}} = Z$. In suspensions where $|L_B Z / \sigma_{\text{Col, Col}}| \gtrsim 1$,
the potential in \expressionname~\eqref{eq:u_DLVO} remains to be a good approximation of $u_{\text{Col, Col}}^{\text{eff}}(r)$
at sufficiently large macroion separation distances, but the effective charge number, $Z_{\text{eff}}$, satisfying $|Z_{\text{eff}}| \leq |Z|$,
can considerably differ from the bare charge $Z$
\cite{Alexander1984, Bitzer1994, Levin1998, Tamashiro1998, Diehl2001, Bocquet2002, Trizac2002, Trizac2003, Trizac2004, Castaneda-Priego2006,
Dobnikar2006, Pianegonda2007, Rojas-Ochoa2008, Torres2008, McPhie2008, Colla2009, Falcon-Gonzalez2010,
Falcon-Gonzalez2011, Castaneda-Priego2012}.

We determine $Z_{\text{eff}}$ in the following by fitting
$u^{\text{DLVO}}_{\text{Col, Col}}(r)$ to
$u^{\text{eff}}_{\text{Col, Col}}(r)$ at large particle separations.
Here, $Z_{\text{eff}}$ is used as the only tunable fit parameter. 
The effective charge $Z_{\text{eff}}$ can be regarded as the overall charge of a colloidal sphere and that part 
of it's surrounding double layer in which the Debye-H\"{u}ckel approximation of microion distributions breaks down.

We note here that our description of the electric double layer is similar, but not equal to the so-called 'Basic Stern Model'
or 'Zeroth-order Stern Model' \cite{Westall1980, Healy1978}. Like these variants of the Stern model, our PM description takes account
of the finite size of microions in using the pairwise additive ion hard-core diameters $\sigma_{i,j}$. However, going beyond the Stern model,
our description also takes account of non-mean-field (PM-HNC) correlations between all ion species,
regardless of the ion separation distance. The Stern model, in contrast, assumes mean-field (Poisson-Boltzmann) microion
distributions in the diffusive (non-condensed) part of the double layer, in the same fashion
as the historically preceding Gouy-Chapman model.

\subsection{Self-consistent solution}\label{sec:sub:sel-cons}

We solve the set of \expressionsname~\eqref{eq:CO2_balance_gammas},~\eqref{eq:water_balance_gammas}, \eqref{eq:Henderson_Hasselbalch}
and \eqref{eq:quadratic_eqn_for_HCO3-}--\eqref{eq:Hansen-Vieillefosse-Belloni} for the eight
unknown quantities in \tablename~\ref{tab:known_and_unknown}, by the iterative
algorithm described in \appendixname~\ref{AppendixB}.
This algorithm seeks a fixed point solution of the coupled set of equations by stepping
repeatedly through the loop of subproblems that is schematically depicted in \figurename~\ref{fig:schematic}.

\section{Results}\label{sec:Results}

\subsection{Ion pair-correlations and pH value}\label{sec:sub:Paircorr}

\begin{figure}
 \includegraphics[width=1.02\columnwidth]{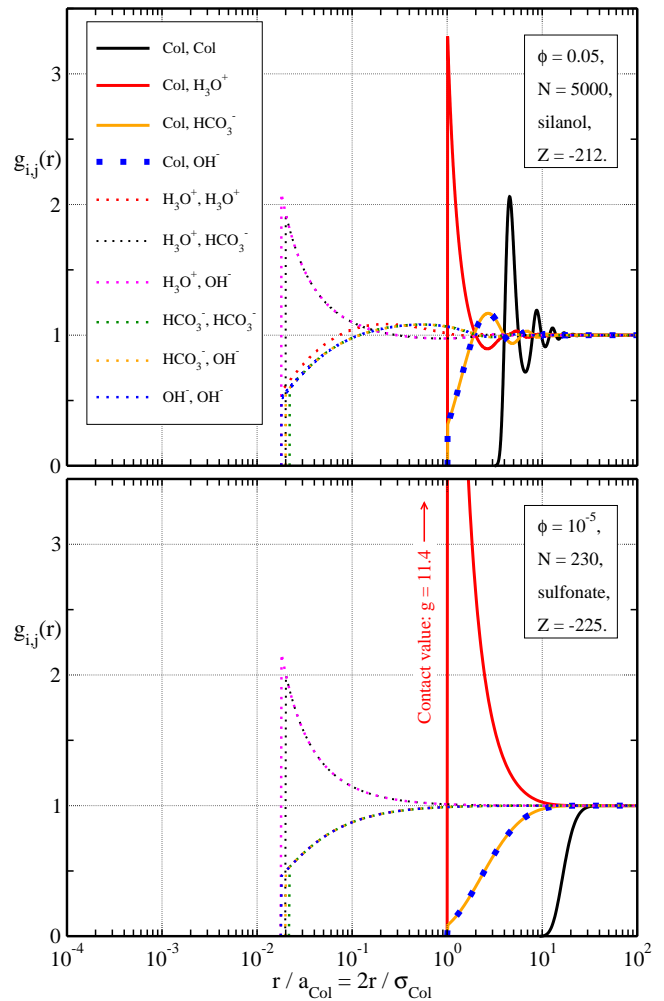} 
 \vspace{-2.5em}
 \caption{(Color online) The PM-HNC partial rdf's $g_{i,j}(r)$ between the four ion species
 Col, H$_3$O$^+$, HCO$_3^-$ and OH$^-$
 (as indicated in the legend), for two different colloidal suspensions.
 Top panel: Suspension at $\phi = 5\%$ colloidal volume fraction, with $N = 5000$ silanol surface groups per colloidal sphere,
 and a resulting colloidal bare charge of $Z = -212$. 
 Bottom panel: Suspension at $\phi = 10^{-5}$, with $N = 230$ sulfonate surface groups per colloidal sphere,
 and a resulting colloidal bare charge of $Z = -225$. 
 The horizontal (logarithmic) and vertical (linear) axes ranges are equal in both panels. In the lower panel,
 $g_{\text{Col, H}_3\text{O}^+}(r)$ (solid red curve) exceeds the vertical axis range.
 The principal maximum of this function is 11.4.
 }
 \label{fig:rdfs}
\end{figure}

As a first result, \figurename~\ref{fig:rdfs} features the PM-HNC solutions for the partial rdf's $g_{i,j}(r)$ between the four ion species
Col, H$_3$O$^+$, HCO$_3^-$ and OH$^-$, in two different colloidal suspensions, corresponding to the two panels of the figure.
The partial rdf's between the ultradilute colloidal sphere species 'dilCol' and other species are indistinguishable from the
corresponding functions for species 'Col', on the scale of \figurename~\ref{fig:rdfs}, and are therefore not shown. 
Results in the top panel of \figurename~\ref{fig:rdfs} are
for a suspension of colloidal spheres that carry $N = 5000$ silanol surface groups.
In the self-consistent solution of the physicochemical problem, only $4\%$ of the silanol surface groups are
dissociated under these conditions, which results in a colloidal bare charge
of $Z = -212$. Due to the strong electrostatic repulsion and the relatively high colloidal volume fraction of $\phi = 5\%$,
the pair correlations between colloidal spheres in this suspension are rather strong, as characterized
by a macroion-macroion rdf principal maximum of $g_{\text{Col, Col}}(r \approx 6a_{\text{Col}}) = 2.06$
(black solid curve in the upper panel of \figurename~\ref{fig:rdfs}). The concentration of positive hydronium ions close to the 
negatively charged colloidal sphere's surfaces is $3.3$ times higher than the suspension-averaged hydronium ion concentration,
as indicated by the contact value $g_{\text{Col, H}_3\text{O}^+}(\sigma_{\text{Col, H}_3\text{O}^+}) \approx 3.3$
(red solid curve in the top panel of \figurename~\ref{fig:rdfs}).

The lower panel of  \figurename~\ref{fig:rdfs} features the partial rdf's for a dilute suspension, at a colloidal volume fraction of $\phi = 10^{-5}$.
Here, each colloidal sphere carries $N = 230$ sulfonate surface groups. Due to the small value of the surface group
acidic dissociation constant, \mbox{$pK_a^{\text{R-O-SO}_3\text{H}} = 1.5$}, the self-consistent solution of the physicochemical set of equations
predicts that $98\%$ of the sulfonate groups are dissociated here, resulting in a colloidal bare charge of $Z = -225$.
Attraction of diffusing hydronium counterions towards the colloidal sphere's surfaces is strong,
as signaled by the contact value, $g_{\text{Col, H}_3\text{O}^+}(\sigma_{\text{Col, H}_3\text{O}^+}) = 11.4$, of the
macroion-counterion rdf (red solid curve in the lower panel of \figurename~\ref{fig:rdfs}).
Noting that the low-density (mean-field) approximation
$g_{\text{Col, H}_3\text{O}^+}(\sigma_{\text{Col, H}_3\text{O}^+}) \approx
\exp\left\lbrace -\beta u_{\text{Col, H}_3\text{O}^+}(\sigma_{\text{Col, H}_3\text{O}^+})\right\rbrace = 22.7$
predicts a contact value that is two times too large, we conclude that the
non-mean-field character of microion distributions is a strong effect that must not be neglected under these conditions.

\begin{figure}
 \includegraphics[width=0.75\columnwidth,angle=-90]{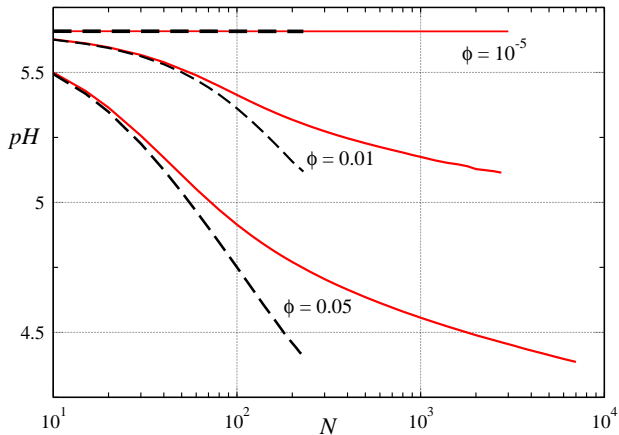} 
 \vspace{-2em}
 \caption{(Color online) Suspension $pH$-values as functions of the number of acidic surface groups, $N$, per colloidal sphere.
 Red solid curves are for weakly acidic silanol surface groups, and
 black dashed curves are for the more strongly acidic sulfonate surface groups.
 Results for three colloidal sphere volume fractions, $\phi = 10^{-5}, 0.01,$ and $0.05$ are shown, as indicated in the figure.}
 \label{fig:pH}
\end{figure}

In \figurename{}\ref{fig:pH}, we display the $pH$-values of different colloidal suspensions, as functions of the number, $N$, of acidic surface
groups per colloidal sphere. Red solid curves are for colloidal spheres with silanol surface groups, and black dashed curves are
for colloidal spheres that carry the more strongly acidic sulfonate surface groups. Results for three different colloidal volume
fractions, $\phi = 10^{-5}, 0.01,$ and $0.05$ are shown in \figurename{}~\ref{fig:pH}. At the lowest volume fraction, $\phi=10^{-5}$,
the $pH$-value is practically independent of $N$. The reason is, that the amount of hydronium ions which are released by the colloidal spheres into suspension
is negligible, compared to the number of hydronium ions created in bulk suspension in the two reactions in \expressionsname~\eqref{eq:CO2_reaction}
and \eqref{eq:water_selfdissociation}. The resulting value $pH = 5.65$ is a reasonable value for demineralized water that is saturated with CO$_2$
under standard atmosphere.
As the colloidal volume fraction is increased to $\phi = 0.01$ and $0.05$, surface-released hydronium ions
lead to appreciable drops in the $pH$-value. For colloids with sulfonate surface groups, the $pH$-value drops more rapidly (as
a function of $N$ or $\phi$) than in case of the weakly acidic silanol groups.

\subsection{Colloidal bare and effective charges}\label{sec:sub:Zeff_Results}

\begin{figure}
 \includegraphics[width=0.77\columnwidth,angle=-90]{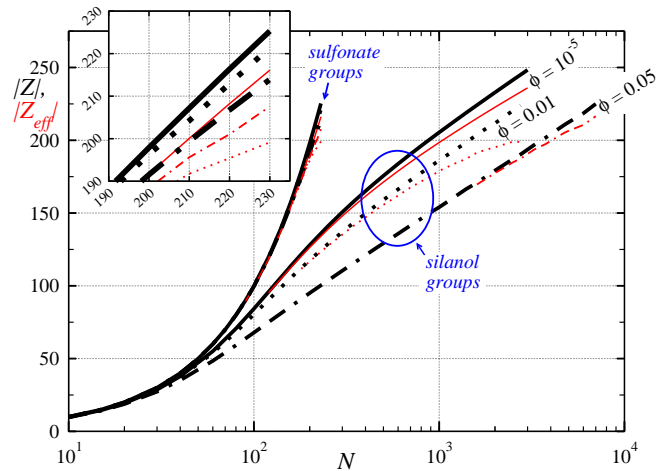}  
 \vspace{-2.5em}
 \caption{(Color online) The absolute values of the colloidal bare charge, $Z$ (thick black curves), and the colloidal
 effective charge, $Z^{\text{eff}}$ (thin red curves) are plotted as functions of the number, $N$, of acidic surface groups per colloidal sphere.
 The six rightmost curves (grouped by the blue ellipse) are for colloidal particles with weakly acidic silanol surface groups, and
 the left group of curves (nearly perfectly overlapping in the main panel with logarithmic horizontal axis) are for
 the more strongly acidic sulfonate surface groups.
 The inset magnifies the details of the sulfonate group results on a linear-linear scale.
 Results for three different colloidal volume fractions $\phi$ are shown:
 Solid curves are for $\phi = 10^{-5}$,
 dotted curves for $\phi = 0.01$, and
 dashed-dotted curves for $\phi = 0.05$.
 }
 \label{fig:Z_Zeff_vs_N}
\end{figure}

In \figurename~\ref{fig:Z_Zeff_vs_N}, we plot the absolute values of $Z$ (black thick curves) and $Z^{\text{eff}}$ (red thin curves),
as functions of the surface group number, $N$.
Once again, the three volume fractions $\phi = 10^{-5}, 0.01$ and $0.05$ are considered.
Solid curves in \figurename~\ref{fig:Z_Zeff_vs_N} are for $\phi=10^{-5}$,
dotted curves are for $\phi=0.01$, and dashed-dotted curves are for $\phi=0.05$.
The six rightmost curves in \figurename~\ref{fig:Z_Zeff_vs_N} (grouped by a blue ellipse)
represent results for colloidal spheres with silanol surface groups.
The six curves on the left side, corresponding to sulfonate surface groups, are nearly overlapping
on the logarithmic-linear scale of the main panel. The end regions of these curves at $N \lesssim 230$
are magnified in the inset, on a linear-linear scale.
As the number, $N$, of surface groups increases, the colloidal bare and effective charges also increase monotonically.
In case of sulfonate surface groups, $Z(N)$ and $Z^{\text{eff}}(N)$ rise more quickly than in case of silanol surface groups.
Nearly all sulfonate groups are dissociated for all probed suspension parameters, resulting in $|Z| \approx N$.
Dissociation of the more weakly acidic silanol groups is considerably weaker, and becomes significantly and
increasingly suppressed at high values of $N$, where the functions $|Z|(N)$ increase only logarithmically.

The ratio $Z^{\text{eff}} / Z$ of colloidal effective and bare charge decreases as a function of $N$,
which is due to an increasing number of microions with non-Debye-H\"{u}ckel like distributions.
In the suspension with $\phi = 0.01$ and $N=2750$ silanol groups per colloidal sphere, only $8\%$ of the surface
groups are dissociated in equilibrium, resulting in $Z = -224$, and nonlinear screening
leads to a further diminished value of the effective charge of $Z^{\text{eff}} = -199$. 

The value of $|Z|$ decreases monotonically when $\phi$ is raised.
This is due to two reasons:
First, the $pH$-value, entering the charge regulation \expressionname~\eqref{eq:Henderson_Hasselbalch}, drops
with increasing $\phi$, starting from its \mbox{CO$_2$-buffer} controlled limit $5.65$
(\textit{c.f.}, \figurename~\ref{fig:pH} and \refname~\cite{Labbez2009}).
Secondly, increasing $\phi$ causes increasing number densities of microions that interact
electrostatically with the acidic surface groups, thereby increasing the excess chemical potential $\ln(\gamma_{\text{Sg$^-$}})$.
Both of these contributions are generally important, reaching similar
magnitudes for the $\phi = 0.05$ silanol surface group system at high values of $N$.

Note that for a fixed value of $Z$, \figurename~\ref{fig:Z_Zeff_vs_N} exposes a non-monotonic dependence of $Z_{\text{eff}}$ on $\phi$:
In case of silanol surface groups and $|Z| = 205$, for example, we find
$|Z_{\text{eff}}| = 198$ at $\phi = 10^{-5}$,
$|Z_{\text{eff}}| = 192$ at $\phi = 0.01$, and 
$|Z_{\text{eff}}| = 200$ at $\phi = 0.05$,
\textit{i.e.}, an initially decreasing $|Z_{\text{eff}}|(\phi)$ which then increases.
The same effect is also observed in case of sulfonate surface groups (see here the inset of \figurename~\ref{fig:Z_Zeff_vs_N}).
In fact, also mean-field effective charge calculations show such behavior \cite{Gisler1994, Trizac2004}.
The observed nonmonotonicity in $|Z_{\text{eff}}|(\phi)$ can be understood as follows:
In the infinite dilution limit, the entropic gain for counterions diffusing in the bulk beats the gain in electrostatic binding energy near the colloidal surfaces.
Hence, all counterions diffuse away from the colloidal sphere surfaces, and $|Z_{\text{eff}}|$ is (nearly) equal to $|Z|$ for $\phi \to 0$.
When $\phi$ is increased, the expected non-Debye-H\"{u}ckel like distribution of microions about the colloidal surfaces sets in, resulting in a decrease of $|Z_{\text{eff}}|$.
When $\phi$ is further increased, global electroneutrality demands that the microion number densities in bulk solvent (\textit{i.e.}, far away from the colloidal sphere's
surfaces) continue to increase, and the result can be a reducing electrostatic energy penalty for a counterion that diffuses
from a colloidal surface into the bulk. In the bulk, the counterion itself experiences now an appreciable screening of its electric field,
caused by the presence of the many other microions. A counterion with a very strongly screened electric field will ultimately behave like an uncharged 
hard sphere and will not condense onto the colloidal surface at all. Therefore, at high $\phi$, $|Z_{\text{eff}}|/|Z|$ can rise again.
\textit{C.f.}, here, the similar effect that has been found in simulations of protein solution at high salinity \cite{Allahyarov2003}.

We finally note, that the present approach is similar in spirit to the determination of effective charges
from elasticity experiments \cite{Lindsay1985}. There the shear modulus of a randomly oriented polycrystalline colloidal
solid is determined and interpreted in terms of an effective DLVO pair potential [\textit{c.f.}, \expressionname~\eqref{eq:u_DLVO}],
with $Z_{\text{eff}}$ as the only free fit parameter. This implies an account for nonlinear screening,
but furthermore also for the so-called macroion shielding effect \cite{Klein2002}, \textit{i.e.},
the screening of the macroion-macroion pair potential due to the presence of other macroions.
Consequently, the effective elasticity charge is lower than any electro-kinetic charge measured on the very same suspension
\cite{Wette2002, Wette2003}. Within a mean-field level description, macro-ion shielding is a many body effect \cite{Brunner2004},
which considerably complicates the search for suitable pair-interactions \cite{Trizac2004, Castaneda-Priego2006}.
It becomes most important, when the range of the repulsion exceeds the nearest neighbor distance,
\textit{i.e.} close to the fluid-solid phase transition. It appears to vanish at strong screening,
or at elevated volume fractions \cite{Shapran2006}. If macroion shielding effects are subsumed under the elasticity effective charge,
the latter can be used to predict, \textit{e.g.}, the fluid-solid phase boundary for this suspension employing the results
of Monte Carlo simulations for charged spheres interacting via a Yukawa-type pair potential \cite{Robbins1988, Wette2006, Wette2010}.
Also in the present approach all electrostatic interactions are accounted for within the PM, which naturally includes
the macroion shielding effect. Our effective charge number $Z_{\text{eff}}$, obtained from mapping the macroion-macroion
effective interaction potential to a DLVO-type pair potential, should therefore yield a suitable input for calculations
of the suspension's fluid structure on the OMF level, and allow predictions for experimentally measurable structure factors.

\section{Outlook}\label{sec:Outlook}

The theoretical scheme presented here can be rather straightforwardly generalized to aqueous colloidal suspensions with added salt
or other kinds of reactive electrolytes.
To this end, the salinity-dependent bulk carbon dioxide concentration can be used \cite{Weiss1974, Millero1995}.

Inclusion of sodium hydroxide \cite{Gisler1994} or pyridine \cite{Yamanaka2004,Shinohara2013}
into the theoretical description would be particularly interesting,
since it has been reported that suspensions of colloidal silica spheres exhibit a phase diagram with 
reentrant fluid-solid-fluid phase sequences, when either the concentration of added base or the concentration of
colloidal spheres is increased \cite{Herlach2010, Wette2010}.
Constructing a closed set of equations and obtaining PM-HNC solutions for all ionic rdf's in a realistic model for a
colloidal suspension with added base will be somewhat more complicated than for the solvent model
discussed in the present paper, due to the larger number of neutral and ionic species that will have to be taken account of.

While we have concentrated on aqueous suspensions in this work,
(variations of) the presented formalism should also be applicable to
the prominent problem of charge regulation in non-aqueous colloidal suspensions
\cite{PhilipseVrij1988, Yethiraj2003, Royall2006, Beunis2012}, which can also exhibit 
unusual phase sequences like crystal-fluid-crystal \cite{Royall2006}.
In non-aqueous suspensions the Bjerrum length is one to two orders of magnitude longer than
in aqueous suspensions, which results in much stronger electrostatic interactions.
As a consequence, tight Bjerrum-pairing of microions occurs \cite{Zwanikken2009, Valeriani2010},
and nontrivial ion correlations are of great importance in the screening of colloidal sphere charges.
Incorporation of non-mean-field like ion distributions in a semi-analytical theoretical framework like
the present one would therefore be desirable in case of non-aqueous media.
Note, however, that in non-aqueous media the mechanisms of
colloidal (chemical) charge regulation are far more complex than the simple dissociation of 
surface groups discussed in our present work. Charging of colloidal spheres in non-aqueous
media can arise from an intricate interplay of preferential surfactant adsorption, micelle formation,
and dissociation of counterions from the colloidal surfaces into the hydrophilic core of micelles
\cite{Morrison1993}. One future extension of the present work should be concerned
with the inclusion of these charging mechanisms into the physicochemical problem set.

\section{Conclusions}\label{sec:Conclusions}

We have demonstrated that a set of chemical association-dissociation balances in the colloidal bulk phase and at the
surfaces of colloidal spheres can be coupled by means of liquid integral equations, and that the resulting set
of physicochemical equations can be efficiently numerically solved.
The theoretical scheme introduced here allows for an \textit{ab initio} calculation of colloidal bare charges $Z$
and effective charges $Z^{\text{eff}}$ for fluid colloidal suspensions in a wide range of suspension parameters. As input to the
theoretical scheme one needs to know only the acidic dissociation constants $pK_a$ of the involved chemically reactive species,
the (effective) sphere diameters of the macroions and of all microions, the colloidal volume fraction, and the number, $N$, of
dissociable acidic surface groups per sphere. Different from $Z$ and $Z_{\text{eff}}$, values for $N$ can be directly and straightforwardly obtained
in titration experiments and are therefore experimentally more easily accessible.

The large macroion to microion size- and charge asymmetries in typical
colloidal suspensions cause a huge numerical burden in any relevant computer simulation of the primitive model.
In contrast to this, the self-consistent numerical solution of the scheme presented here takes only few minutes or less on an
inexpensive personal computer, for a given set of suspension parameters. Our method is therefore well-suited for planning and analyzing
experiments with charged colloidal suspensions, and to calculate primitive model pair-correlation
input for theories of colloidal dynamics including electrophoresis, colloidal diffusion and rheology.

\section*{Acknowledgement}
It is our pleasure to thank the anonymous reviewer for helpful suggestions which have improved this paper.
M.H. and H.L. acknowledge funding by the European Research Council (ERC) Advanced Grant INTERCOCOS, FP7 Ref.-Nr. 267499.
T.P. acknowledges funding by the Deutsche Forschungsgemeinschaft (DFG), within the projects SPP1296, Pa 459/16 and Pa 459/17.

\appendix

\section{Surface group chemical potential}\label{AppendixA}

In the PM, colloidal particles are approximated as dielectric hard spheres with solvent dielectric constant $\epsilon$,
and the electric charge is assumed to be homogeneously smeared out on the colloidal sphere's surfaces. Within this model,
which neglects surface-charge patchiness, a monovalent charged surface group represents nothing else than a single elementary
charge that is smeared out about the surface of it's associated colloidal sphere. This allows us to construct an approximate method
to determine the charged surface group excess chemical potential in consistence with the already made PM assumptions. The three-step
method consists of a colloidal sphere extraction step, a charging step, and a colloidal sphere re-insertion step, as described in the
following. In order to keep the suspension globally electroneutral at all steps, hydronium (H$_3$O$^+$) counterions are taken into account.

\begin{center}
\emph{Step 1 (colloidal sphere extraction):}
\end{center}
From the five-component PM ionic suspension described in \sectionname~\ref{sec:sub:HNC}, one colloidal sphere with charge $Ze$ is
extracted and placed into pure solvent, \textit{i.e.}, into an infinite, otherwise particle-free, dielectric continuum with dielectric constant $\epsilon$.
To restore charge neutrality of the suspension, a number of $|Z|$ hydronium counterions are also extracted from the suspension into pure
solvent (and into infinite mutual distance). The change in Gibbs free energy in step one is thus
\begin{equation}\label{eq:DeltaG_step1}
\Delta G_1 = -\mu_{\text{Col}} - |Z|\mu_{\text{H$_3$O$^+$}}. 
\end{equation}

\begin{center}
\emph{Step 2 (charging of the sphere):}
\end{center}
Inside pure solvent, one elementary charge is removed from the colloidal sphere and placed into infinite distance from the sphere.
Then, the removed charge is compressed to the hydronium ion diameter $\sigma_{\text{H$_3$O$^+$}}$. The change in Gibbs free energy in this step is
equal to the change in Coulomb (self-)energy of the electric charge density: 
\begin{eqnarray}
\Delta G_2 &=&
\frac{\displaystyle{2 (Z-1)^2 e^2}}{\displaystyle{\epsilon \sigma_{\text{Col}}}} -
\frac{\displaystyle{2 Z^2 e^2}}{\displaystyle{\epsilon \sigma_{\text{Col}}}} +
\frac{\displaystyle{2 e^2}}{\displaystyle{\epsilon \sigma_{\text{H$_3$O$^+$}}}}.\nonumber\\
&=& \frac{\displaystyle{2 L_B}}{\displaystyle{\sigma_{\text{Col}}}} (1-2Z) k_B T +
\frac{\displaystyle{2 L_B}}{\displaystyle{\sigma_{\text{H$_3$O$^+$}}}} k_B T. \label{eq:DeltaG_step2}
\end{eqnarray}
Step two leaves us with a colloidal sphere of charge $(Z-1)e$ and $|Z-1|$ hydronium ions in pure solvent. 

\begin{center}
\emph{Step 3 (colloidal sphere re-insertion):}
\end{center}
Insert the colloidal sphere of charge $(Z-1)e$ and the $|Z-1|$ hydronium ions from pure solvent into the five-component PM suspension.
The change in Gibbs free energy in this step is:
\begin{equation}\label{eq:DeltaG_step3}
\Delta G_3 = \mu_{\text{dilCol}} + |Z-1|\mu_{\text{H$_3$O$^+$}}, 
\end{equation}
where the index 'dilCol' stands for the ultradilute species of colloidal spheres with charge $(Z-1)e$ each.

Note that, in the thermodynamic limit, none of the five ion number densities in the suspension is changed when steps 1$-$3 are applied.
Therefore, the hydronium ion chemical potentials in step 1 and 3 are exactly equal, and we gain the expression
\begin{eqnarray}
\beta \Delta G
&=& \beta \left[ \Delta G_1 + \Delta G_2 + \Delta G_3 \right]\nonumber\\  
&=& \beta \mu_{\text{dilCol}} - \beta \mu_{\text{Col}} + 
\frac{\displaystyle{2 L_B}}{\displaystyle{\sigma_{\text{Col}}}} (1-2Z) +\nonumber\\
&&\beta \mu_{\text{H$_3$O$^+$}} +
\frac{\displaystyle{2 L_B}}{\displaystyle{\sigma_{\text{H$_3$O$^+$}}}}\label{eq:DeltaG_total}
\end{eqnarray}
for the total change in normalized Gibbs free energy.
The second and third row in \expressionname~\eqref{eq:DeltaG_total} account for the insertion
of a charged surface group and a hydronium ion, respectively. Considering the excess part of all
quantities in \expressionname~\eqref{eq:DeltaG_total}, we thus arrive at the expression
\begin{equation}\label{eq:surfgroup_exc_chempot}
\ln(\gamma_{\text{Sg$^-$}}) = \beta \mu_{\text{Sg$^-$}}^{\text{exc}} = 
\beta \mu_{\text{dilCol}}^{\text{exc}} - \beta \mu_{\text{Col}}^{\text{exc}} + 
\frac{\displaystyle{2 L_B}}{\displaystyle{\sigma_{\text{Col}}}}(1-2Z)
\end{equation}
for the charged surface group activity coefficient $\gamma_{\text{Sg$^-$}}$, which is required as input to the Henderson-Hasselbalch
\expressionname~\eqref{eq:Henderson_Hasselbalch} for the colloidal surface charge.

\section{Iterative self-consistent solution}\label{AppendixB}

Here we present our iterative algorithm for solving the set of
\expressionsname~\eqref{eq:CO2_balance_gammas},~\eqref{eq:water_balance_gammas}, \eqref{eq:Henderson_Hasselbalch}
and \eqref{eq:quadratic_eqn_for_HCO3-}--\eqref{eq:Hansen-Vieillefosse-Belloni} for the eight
unknown quantities in \tablename~\ref{tab:known_and_unknown}:

\begin{center}
\emph{Initialization:}
\end{center}
Choose a colloidal sphere number density [Col], and a fixed number, $N$,
of acidic surface groups per colloidal sphere.
Choose $\text{[H}_2\text{O]} = 54.2$ M and $\text{[CO}_2\text{]} = 1.52 \times 10^{-5}$ M, and
a concentration [dilCol]$\lesssim 10^{-6} \times$ [Col]. 
Initialize the colloid charge number by setting $Z = -N$,
and initialize the thermodynamic activity coefficients by choosing $\gamma_i = 1$ for all ionic species $i$.

\begin{center}
\emph{Step 1:}
\end{center}
Calculate [H$_3$O$^+$] by solving \expressionname~\eqref{eq:quadratic_eqn_for_HCO3-}
with input
$Z$, [Col], [CO$_2$], [H$_2$O], $K$, $K_a^{\text{CO}_2}$, $\gamma_{\text{HCO$_3^-$}}$, $\gamma_{\text{OH$^-$}}$ and $\gamma_{\text{H$_3$O$^+$}}$.

\begin{center}
\emph{Step 2:}
\end{center}
Solve \expressionsname~\eqref{eq:CO2_balance_gammas} and \eqref{eq:water_balance_gammas} for [HCO$_3^-$] and [OH$^-$], respectively, with input
[H$_3$O$^+$], [CO$_2$], [H$_2$O], $K$, $K_a^{\text{CO}_2}$, $\gamma_{\text{HCO$_3^-$}}$, $\gamma_{\text{OH$^-$}}$ and $\gamma_{\text{H$_3$O$^+$}}$.

\begin{center}
\emph{Step 3:}
\end{center}
Calculate $Z$ from \expressionname~\eqref{eq:Henderson_Hasselbalch}, with input
$N$, [H$_3$O$^+$], $K_a^{\text{SgH}}$, $\gamma_{\text{Sg$^-$}}$, and $\gamma_{\text{H$_3$O$^+$}}$.

\begin{center}
\emph{Step 4:}
\end{center}
Solve the HNC-scheme \expressionsname~\eqref{eq:O-Z}-\eqref{eq:pair_pot} with input
$Z$, [Col], [dilCol], [H$_3$O$^+$], [HCO$_3^-$], and [OH$^-$],
by means of the algorithm from \refname{}~\cite{Heinen2013}.
Then, compute the activity coefficients
$\gamma_{\text{H$_3$O$^+$}}$, $\gamma_{\text{OH$^-$}}$, $\gamma_{\text{HCO$_3^-$}}$, and $\gamma_{\text{Sg$^-$}}$
from \expressionsname~\eqref{eq:Hansen-Vieillefosse-Belloni} and \eqref{eq:surfgroup_exc_chempot}.
Continue with step 1.\\

The iteration is stopped once that the relative change in the obtained value of $Z$
is less than $10^{-4}$ in two subsequent loop iterations.

Improved numerical stability is achieved if $Z$ is multiplied by a damping factor at early iteration stages.
The damping factor should be picked from the interval $(0,1]$, and should gradually approach unity during the first few iterations.  
Numerical stability can be further increased if the new solution for $Z$ in step 3 is mixed with the
previous value in proportions $\alpha$ and $(1-\alpha)$, with a mixing coefficient $0 < \alpha < 1$.

\end{document}